# Fractal Growth of Rotating DLA-clusters

A.Loskutov[*], D.Andrievsky, V.Ivanov, K.Vasiliev, A.Ryabov

Physics Faculty, Moscow State University, 119899 Moscow, Russia.

SUMMARY: A theoretical model for fractal growth of DLA-clusters in two- and three-dimensional Euclidean space is proposed. This model allows to study some statistical properties of growing clusters in two different situations: in the static case (the cluster is fixed), and in the case when the growing structure has a nonzero rotation around its germ. By the direct computer simulation the growth of rotating clusters is investigated. The fractal dimension of such clusters as a function of the rotation velocity is found. It is shown that for small enough velocities the fractal dimension is growing, but then, with increasing rotation velocity, it tends to the unity.

## Introduction

In this study we consider a problem of fractal growth via adsorption in the plane and in three-dimensional Euclidean space. The characteristic property of such processes is that the fractal growth is not described by the known equations; *statistical models* are usually used. The most known process of irreversible adsorption is the *diffusion limited aggregation* (DLA) [1-4]. In the simplest case this is a cluster formation when the randomly moving particles are sequentially and irreversibly deposited on the cluster surface. As a result of such clusterization a DLA-fractal is formed. Depending on the embedding dimension $D_e$ the fractal dimension $D_f$ (the exponent in the dependence of the “mass” of the DLA-cluster on the characteristic linear size $M \sim R^{D_f}$) takes the following values [5-6]:

**Table 1.**

| $D_e$ | $D_f$ | $D_e$ | $D_f$ | $D_e$ | $D_f$ | $D_e$ | $D_f$ |
|---|---|---|---|---|---|---|---|
| 2 | 1.71 | 3 | 2.49 | 4 | 3.40 | 5 | 4.33 |

These values are obtained by direct numerical simulations, and for $D_e>5$ there are no exact data. At the same time, analytical investigations are in a quite satisfactory agreement with numerical simulations, but only for $D_e$=3,...,5.

During two last decades the fractal properties of growing DLA-clusters have been widely discussed in the scientific literature [1, 2]. The enhanced interest to this model is connected with a variety of its applications to description of different physical phenomena. However, a very interesting extension of the problem, namely the study of DLA-fractals which are growing under rotation has not been investigated in detail. We present here both the theoretical approach which enables to calculate the fractal dimension of rotating DLA-clusters and the results of direct computer simulation.

We propose a statistical annular model of the fractal growth in $D_e$=2 and $D_e$=3 Euclidean spaces ($D_e$>3 to be obtained). This model allows us to find the dimension of fixed DLA-clusters ("classical" DLA) and the dimension of clusters which have a rotation around the nucleating center. Therewith the convergence of numerical algorithms is two order magnitude faster than for usual case of DLA.

We have performed also direct computer simulation of DLA under rotation. Our preliminary results indicate the transition from fractal to non-fractal behavior of growing cluster on different length scales upon increasing the rotation velocity.

## Theoretical Model

Suppose that particles have a diameter *d*. Let us divide the corresponding embedding space into $n_{\max}$ concentric rings with a middle radius $r_n = nd, n = 0,...,n_{\max}$. Assume that the center of rings coincides with the center of a fractal, and the width of rings is equal to *d*. Let us consider models with the following properties: *a*) particles can be situated only within some ring and can jump from one ring to another; *b1*) within a ring $N_n$ particles can occupy $M_n = [2\pi r_n / d]$ cells, where [...] is an integer part; *b2*) particles can be arbitrary distributed within the ring. The first case *b1* corresponds to cellular model. The second case (*b2*) is transitive between continuous and cellular models. When a particle jumps to another ring, it can be adsorbed here with probability corresponding to collision with one of *N* particles in the ring. It is obvious that in the case of cellular model this probability is

$$\tilde{p}_n = \frac{N_n}{M_n}, \qquad (1)$$

where $N_n$ is the number of particles in an *n*-th layer (ring). In the case of equidistributed particles the adsorption probability is the following:

$$\tilde{p}_n = 1 - \frac{(L_n - (N_n + 1)d)^{N_n - 1}}{(L_n - d)(L_n - (N_n + 1)d)^{N_n - 2}} \theta[L_n - (N_n + 1)d], \qquad (2)$$

where $L_n = 2\pi r_n$ is the length of a circle along which particles are distributed, and $\theta[x]$ is a Heaviside function. Calculations of this probability is a quite simplest task of the known random sequential adsorption (RSA) theory: $N$ particles of a length $d$ are arbitrary distributed on a segment of the length $L$. We throw another particle on this segment; What the probability that this particle will not cover any other one?

Suppose that the particle is within $n$-th layer. Then the jump probability to ($n$+1)-th layer is $p_n^1 = (1 + d/r_n)/3$; to ($n$-1)-th layer is $p_n^{-1} = (1 - d/r_n)/3$, and the probability to be in the same layer is $p_n^0 = 1/3$. In general, $p_n^k = (1 + kd/r_n)/3$, $\sum_k p_n^k = 1$, k=-1, 0, 1. The term $\pm d/r_n$ follows from the assumption that the jump probability to a ring should be proportional to its length. Now we can obtain the probability of adsorption of the particle in the $n$-th layer:

$$P_n = \sum_{k=-1}^{1} \tilde{p}_{n+k} p_n^k. \qquad (3)$$

Here $\tilde{p}_{n+k}$ is the probability of adsorption on the particles within ($n$+$k$)-th layer and $p_n^k$ is the jump probability to this layer. If the adsorption does not take place, the particles moves to an other layer with the probability $Q_n = 1 - P_n$. Corresponding jump probabilities under the condition that the adsorption did not take place, are the following:

$$q_n^k = \frac{p_n^k (1 - \tilde{p}_{n+k})}{Q_n}. \qquad (4)$$

These expressions follow from Bayes formula of the full probability.

Thus, the constructed fractal is a set of layers which are characterized by a pair of numbers $M_n$ and $N_n$. For the simplicity, let us assume that the particle mass is equal to the unit. Then the fractal mass will be equal to $m = \sum_{n=1}^{n_{\max}} N_n$. The dimension of such a fractal can be obtained from the dependence $m(n_{\max})$. In the real DLA process we get $m(n_{\max}) \propto n_{\max}^{D_f}$, where $D_f$ is a fractal dimension. In the logarithmic scale this dependence is a straight line with the slope coefficient $D_f$. Additionally, we can define the local dimension as a function of $n_{\max}$,

$$D_f(n_{\max}) = \frac{d \log m(n_{\max})}{d \log n_{\max}}. \quad (5)$$

The constructed model can be easily generalized for a 3D fractal. In this case the jump probabilities are equal to

$$p_n^k = 1 + k\frac{d}{r_n^2}, \quad k = -1,0,1 \ .$$

The maximal number of particles in a layer can be estimated as follows: $M_n = \left[\pi r_n^2 / d^2\right]$. The adsorption probability (see (1)) in the cellular model is the same, but in the case of an arbitrary distribution of particles there is no any exact expression. However, we can construct many approximations. At the same time, expressions (3) and (4) are the same as for the 3D fractal.

Now let us consider a rotating fractal. As it follows from direct numerical simulations (see below), in this case one or two growing branches of the fractal are predominate, and particles are distributed within them in a quite compact way. So, the assumption that particles are equidistributed within a layer is not correct. Therefore we should consider the cellular model.

Let us denote a frequency of the fractal rotation by ω. We propose the following model: rotation is a continuous process, but particles can make their jumps at a time interval τ. Between jumps the particle is at rest, and can be adsorbed at any time. Let us pass to a rotational frame of reference. Then during time τ the particle is shifted to an angle $\varphi = \omega\tau$ . In this case we have the effective particle size:

$$d_{ef} = d + \varphi r_n = d + \omega\tau r_n .$$

In addition, suppose that the layer already contains $N_n$ particles. Then the adsorption probability is

$$\tilde{p}_n = \frac{N_n + \omega\tau r_n / d}{M_n}.$$

The other expressions remain the same.

## Numerical Investigations of the Model

The model described above has been numerically investigated in two following cases: in 2D case as a cellular model of packing particles in layers with and without rotation, and in 3D case. In each case the logarithmic dependence of the total mass of particles inside a ring of radius *r* on the value of *r* was obtained. The fractal dimensions were calculated by means of a linear approximation (as tangent of the slope angle). In our numerical simulations we have supposed that the diameter of particles is $d = 1$, the jump interval from one layer to another is $\tau = 1$, and the mass of a particle is equal to the unity. So, the ring radius is $r_n = n$. The fractal dimension as a function of the fractal size was calculated by means of the numerical differentiation of curves shown in Fig.1 (see Fig.2 and (5))

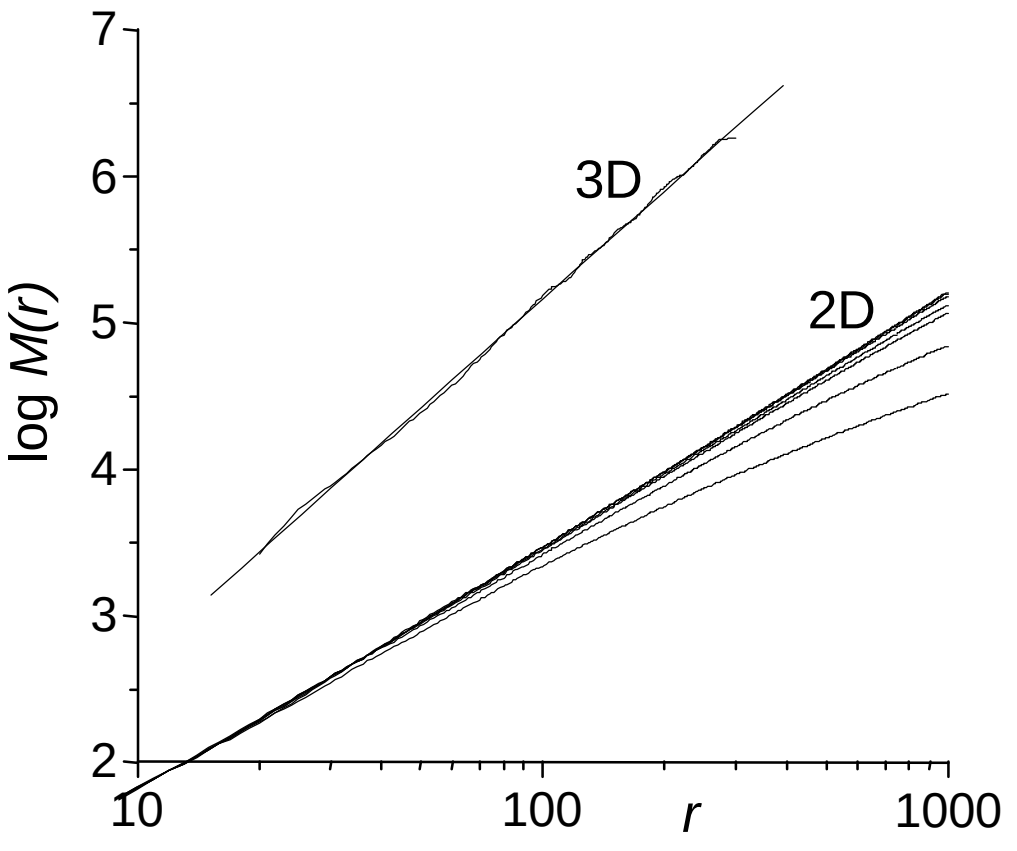

**Fig. 1.** The mass of the fractal as its function of its radius for 2D and 3D spaces

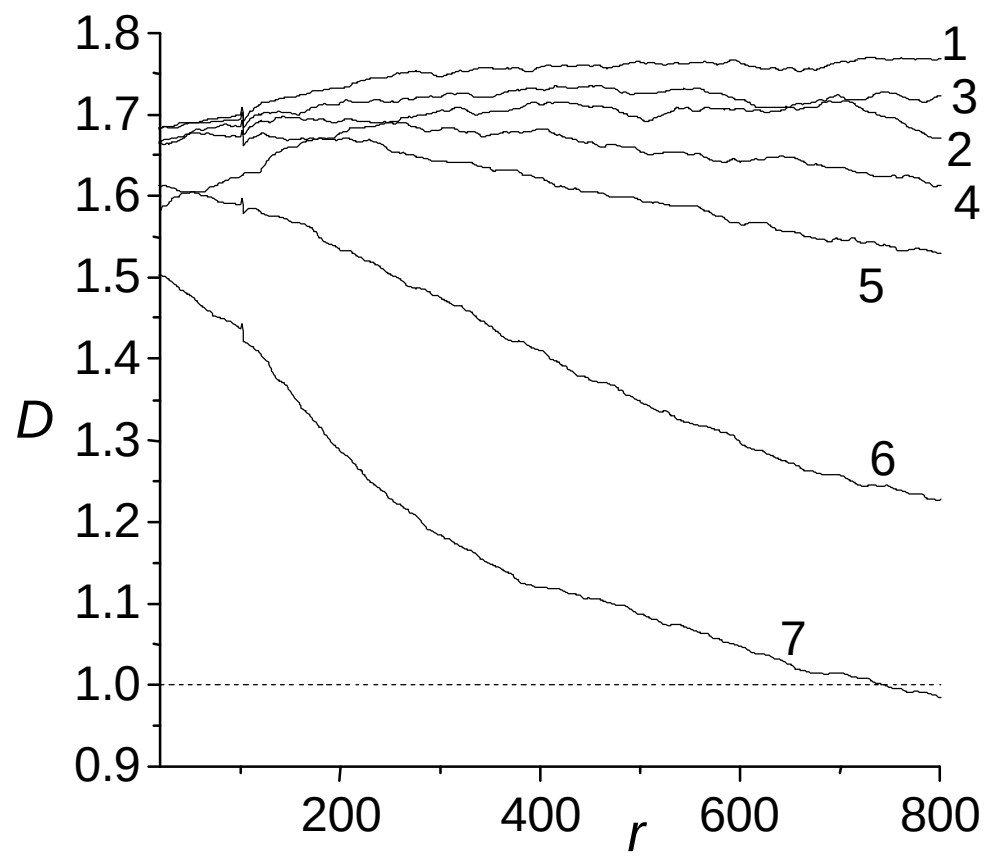

**Fig. 2.** Dependence of the fractal dimension on the fractal radius for different rotating frequencies

Fig.1 shows the dependence of logarithm of fractal mass as a function of radius for 2D and 3D fractals. In the 2D case the set of graphs corresponds to different rotation frequencies (the values are shown below). The smallest slope of the graph corresponds to the maximal frequency. We have obtained 100 fractals for every rotation frequency, and their mean value has been calculated. The visual distinction of graphs is not so radical, however the regression analysis allows us to find some differences. The fractal dimensions calculated by means of the linear approximation are shown in the following Table 2:

**Table 2**

| | 1 | 2 | 3 | 4 | 5 | 6 | 7 |
|---|---|---|---|---|---|---|---|
| $\omega \cdot 10^4$ | 0 | 0.3 | 1 | 3 | 5 | 15 | 45 |
| $D_f$ | 1.727 ± 0.002 | 1.722 ± 0.002 | 1.709 ± 0.002 | 1.675 ± 0.002 | 1.644 ± 0.002 | 1.504 ± 0.003 | 1.303 ± 0.005 |

As it follows from numerical simulations, the fractal dimension decreases when the frequency growths. The results of numerical differentiation of graphs in Fig.1 are shown in Fig.2. This analysis is more detailed. The slope of curves (dimension) is varied in some limits. In the case of the rotating fractals the dimension tends to the unity when the radius is growing. The numbers in Fig. 2 corresponds to columns in Table 2. They mark curves in accordance with different frequencies. For example, number 1 corresponds to a fractal without rotation.

In the 3D case the linear approximation of graphs (see Fig.1) gives the following value of fractal dimension: $D_f$ = 2.471 ± 0.005. This result is in a quite good agreement with the direct numerical simulation (see below).

As it follows from the obtained results, for the rest fractals the statistical approach to fractal growth is in a fine coincident with the results of direct numerical simulations of DLA. Obviously, the differences in the dimension obtained in the case of *rotating* fractals are the results of different methods of the calculation. As in the case of any statistical approach, the proposed model of fractal growth allows us to get only *global* properties of the system, while the use of the sliding windows method supposes knowledge of information about *local* properties. Substantively, the method of sliding windows allows to estimate the dimension of a fractal branch, whereas for the statistical approach there are no the notion of the fractal branch., and the dimension is calculated for the whole fractal set. Moreover, application of the statistical approach assumes implicitly an isotropic particle distribution. Evidently, appearance of anisotropy in the rotating fractals (i.e. in the case of prevalence of some branch) can explain increasing the fractal dimension.

The interesting and but not yet investigated questions are the following: at what fractal radius $r_{cr}$ a certain fractal branch begins to dominate? It is obvious that $r_{cr} = r(\omega)$. What is the growing rate of a fractal and what is the law of twisting of fractal branches?

## Direct Computer Simulation of Rotating DLA-clusters

We consider the DLA-fractal growth in two-dimensional space on the squared lattice. The size of simulation box was equal to 4000×4000 lattice sites. A diffusing particle starts at the randomly chosen point on the circle of radius $1.5R_{cl}$ centered at the origin of coordinate system where the cluster origin is placed. Here, $R_{cl}$ is the maximal distance from the cluster origin to the occupied sites. The particle diffuses freely inside the square box with the side length of $2R_{cl}$ and with rejecting walls until it is added to the growing cluster. As usually this happens when the particle visits the lattice site which is one of the nearest neighbors to one of the occupied sites. In this moment a new diffusing particle begins its random walk. A snapshot of a usual DLA-cluster is presented in Figure 3.

To model the rotation of the DLA cluster we have added a constant angular velocity to the diffusion motion of a particle. This means that it is necessary to add to a random displacement of the particle on the lattice also a constant displacement along the circle centered at the origin of coordinate system and with the radius which is equal to the current distance of diffusing particle from this origin. The particle has usually non-integer coordinates $(x_f, y_f)$ which are transformed into the integer ones $(x,y)$ by taking the integer part: $x = int(x_f)$, $y = int(y_f)$. The angular velocity ω was kept constant. The value $\omega=1$ corresponds to the rotation of the particle on the angle $d\varphi$ at each time step, where $d\varphi$ is chosen in such a way that the particle located at the maximal possible distance from the cluster origin could not jump over one lattice site. Each cluster grows until $R_{cl} < R_s$, where $R_s$ is the maximal size of the lattice, $R_s=1000$, or until its mass is less than 50000 occupied sites. According to maximal possible size of the cluster we have chosen $d\varphi=1/R_s=0.001$ *rad.* Our angular velocity ω is measured in radian per one Monte Carlo step.

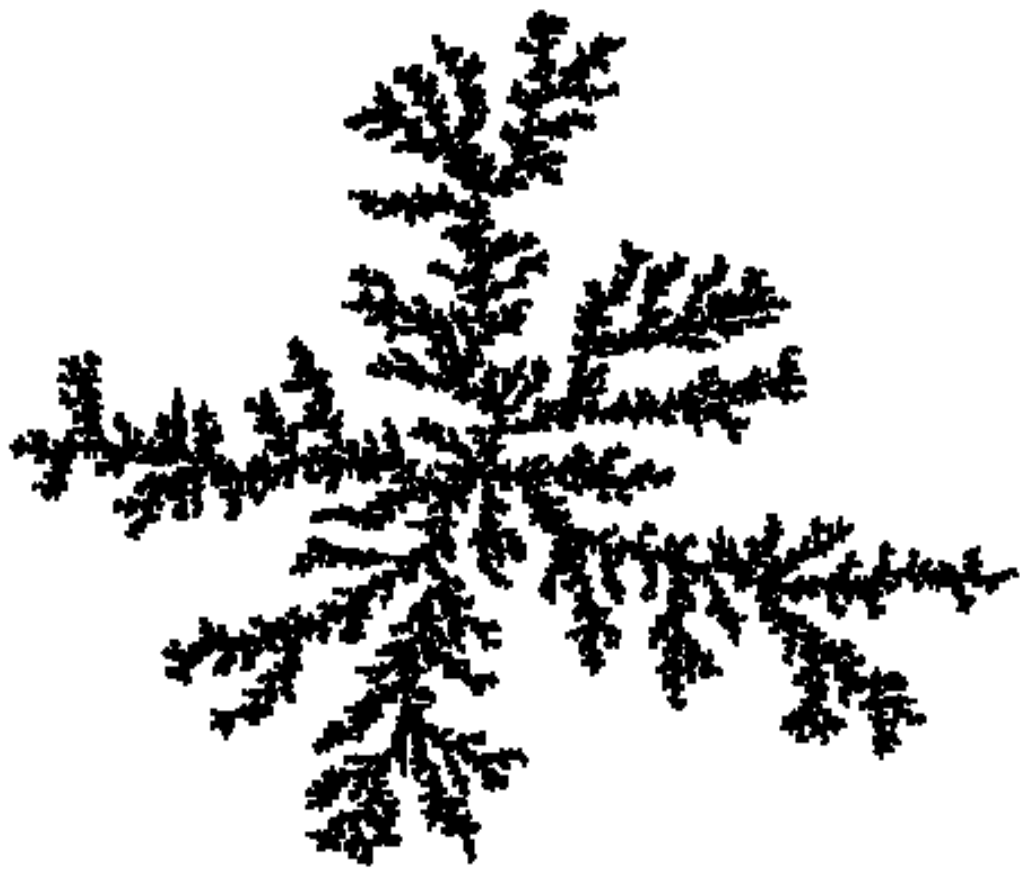

**Fig. 3.** DLA cluster without rotation (M = 50000)

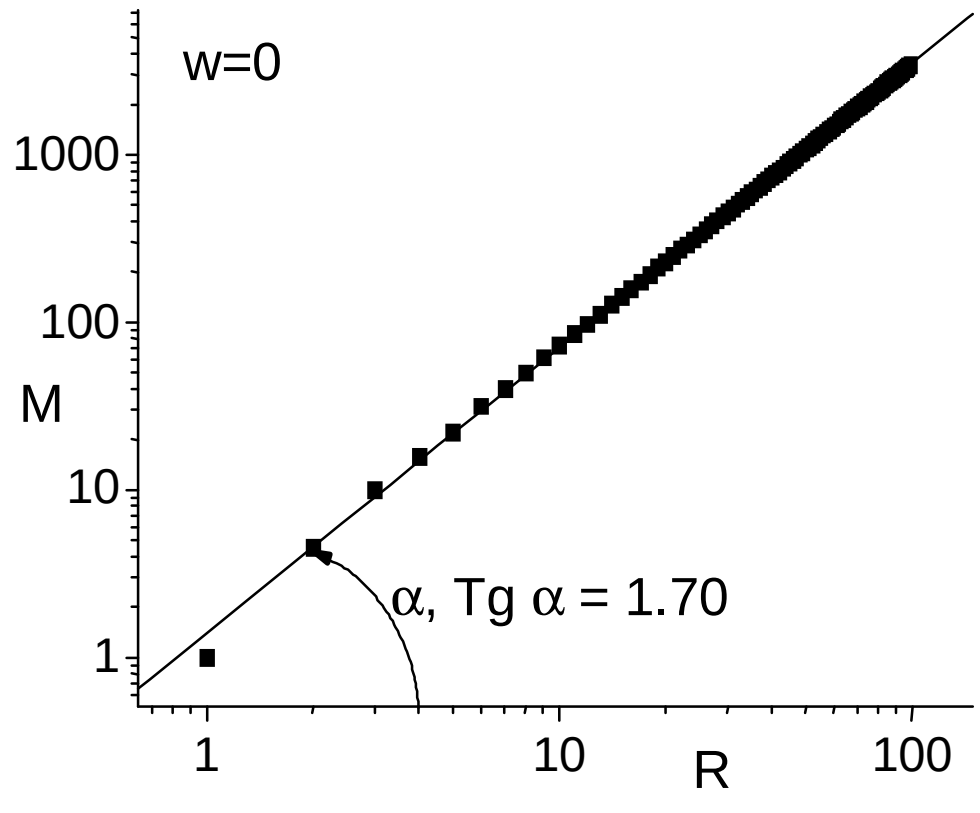

**Fig. 4.** Mass vs. Radius dependence for usual DLA cluster

For each value of the angular velocity we sampled the statistics of 10 independent DLA-clusters. We have calculated the fractal dimension of DLA-aggregates by analyzing the cluster mass as a function of its radius $M \propto R^{D_f}$, where $M$ is the number of occupied sites in the circle $R$, and $D_f$ is the fractal dimension of DLA-cluster. This dependence is presented in Figure 4 (log-log plot) for the case of usual DLA without rotation. To improve the statistics the box counting method ("sliding windows") has been used: 100 points have been chosen randomly on the cluster which were treated as the center of a square with the side *R* from 1 to 100 and the number of occupied sites inside that square was calculated. The dependence *M(R)* was averaged over those 100 different "window" centers and over 10 independent clusters.

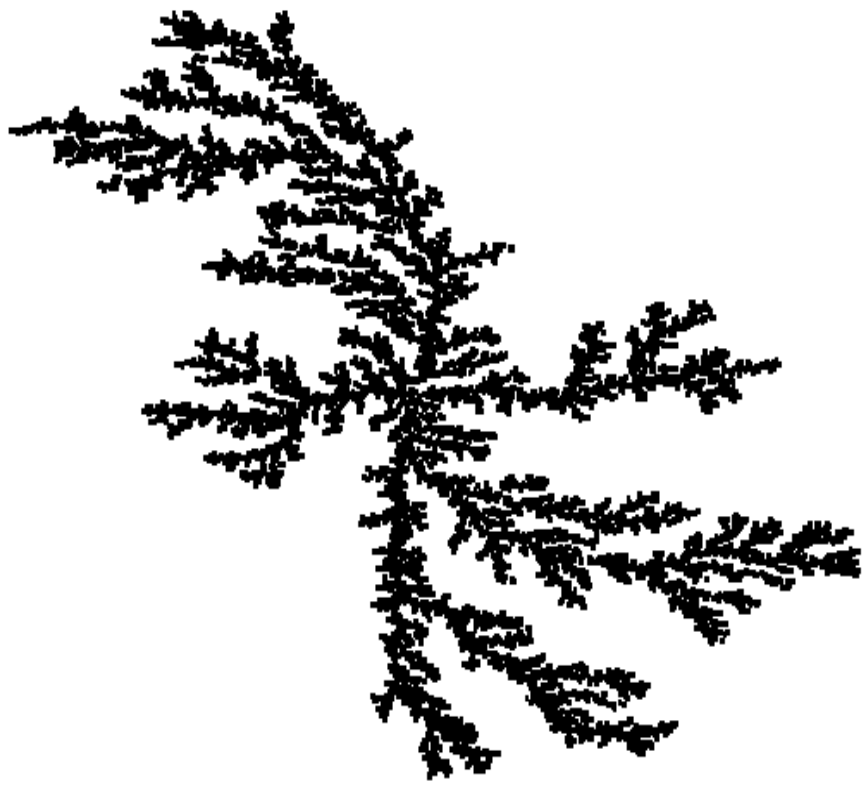

**Fig. 5.** DLA cluster with rotation: $\omega = 0.03 \cdot 10^{-3}$, M = 50000

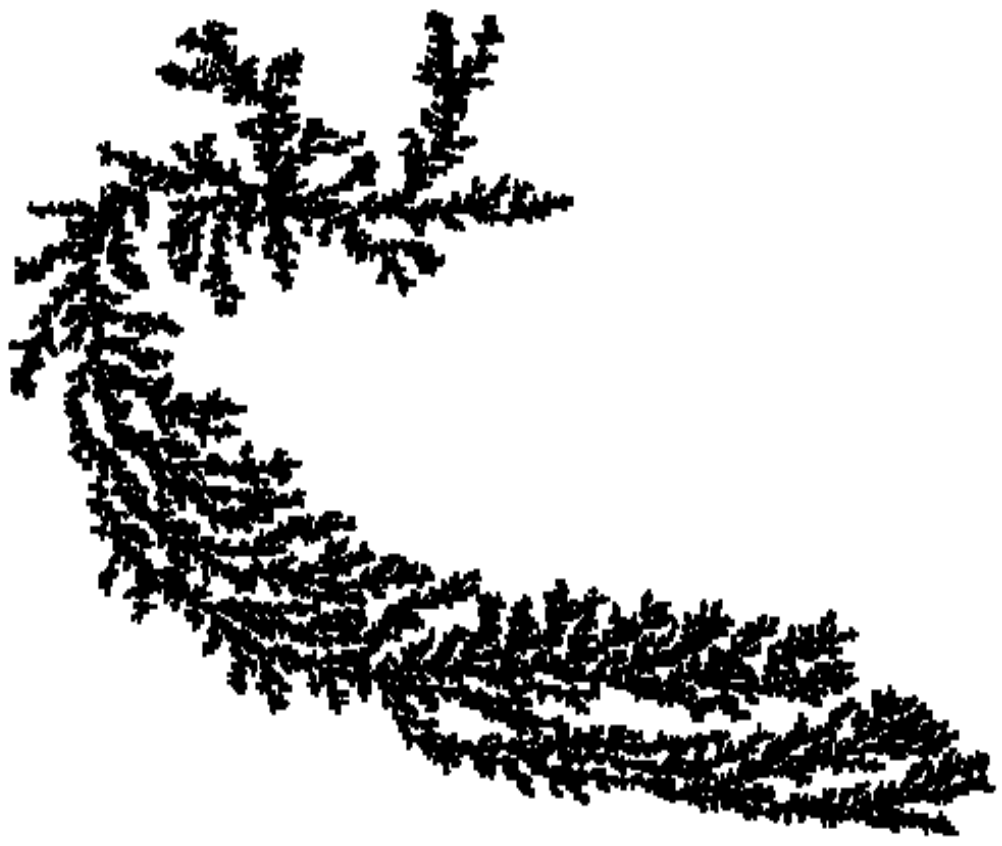

**Fig. 6.** DLA cluster with rotation: $\omega = 0.1 \cdot 10^{-3}$, M = 50000

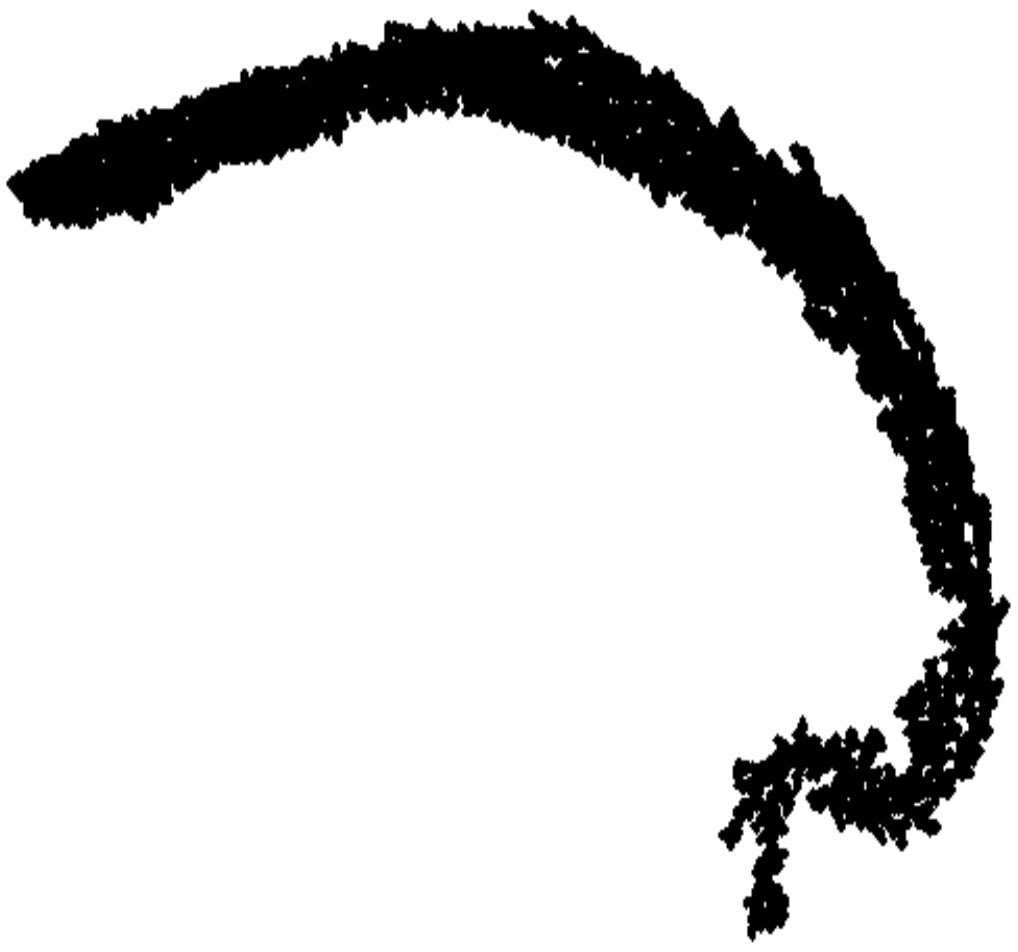

**Fig. 7.** DLA cluster with rotation: $\omega = 0.5 \cdot 10^{-3}$, M = 50000

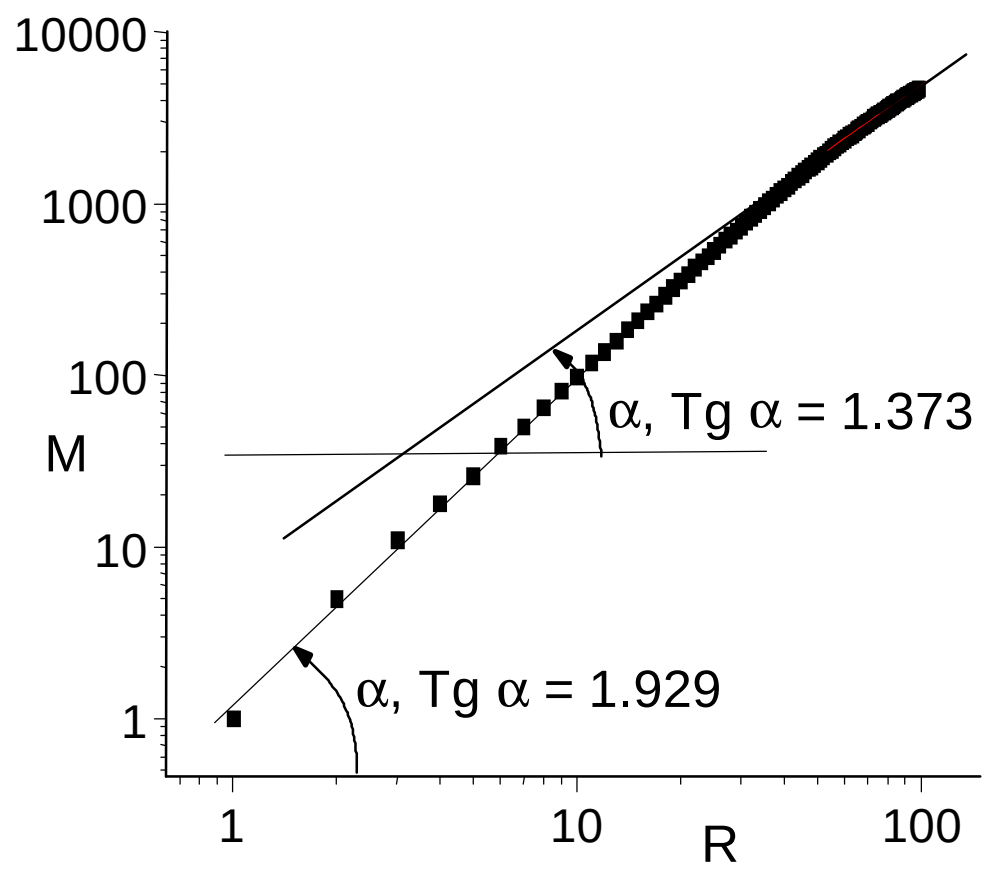

**Fig. 8.** Mass vs. Radius dependence for rotating DLA cluster. $\omega = 0.5 \cdot 10^{-3}$

On Figures 3, 5, 6, 7, 9 the DLA clusters for different angular velocities ω are shown. One can clearly see how clusters are changed upon increasing ω. At ω=0 the cluster is practically isotropic (Fig.3). First, upon increasing ω two "arms" become to be more pronounced (Fig. 5). Then only one of these two "arms" begins to dominate, and finally, the cluster practically degenerates into a "tail" (Fig. 6). At the further increasing ω this "tail" becomes more thin (Fig. 7) and tends to a thin line (Fig. 9). It is possible to indicate several regimes in the cluster growth. If $R_{cl}<R_1(\omega)$ then several "arms" are growing; after this, for $R_1<R_{cl}<R_2$ one "arm" starts to dominate, and, finally, for large enough $R_{cl}>R_2$ this "arm" changes its curvature as follows from Fig.7.

On Figures 4, 8, 10 for different angular velocities of the DLA-cluster rotation from ω=0 to $4.5 \cdot 10^{-3}$ the “mass” of a DLA fractal as a function of radius of a window is presented (double logarithmic plots; the slope is equal to the fractal dimension $D_f$). Starting from some rotation velocity ($\omega \approx 0.3 \cdot 10^{-3}$) there is a clear bending point on the plot (Fig. 8; regions I and II). This can be explained by the fact that starting from this velocity the width of the “tail” becomes smaller than the maximal size of “sliding windows” which is equal to 100. Therefore by measuring the number of occupied sites the empty space inside the window starts to play important role. A position of the bending point indicates the width of the “tail”. Upon increasing ω the slope of the line in the region I goes to 2, and the slope in the region II tends to the unity. The cluster becomes more dense and degenerates into the thin stripe or even into a line. The bending point is shifted to the region of small radii (the width of the stripe is decreasing). At $\omega > 1.5 \cdot 10^{-3}$ one can find even three regions on the plot (Fig. 10; regions I, II and III). It turns out that the slope in the region III is smaller than the unity. This can be explained only as a finite size effect, namely, the size of the window is large enough, and if the window is placed close to the both ends of the thin line it includes too much empty space. We are studying now the finite size effects more attentively. On Fig.11 the dependencies of the slopes in the regions I, II и III are presented as functions of the rotation velocity ω.

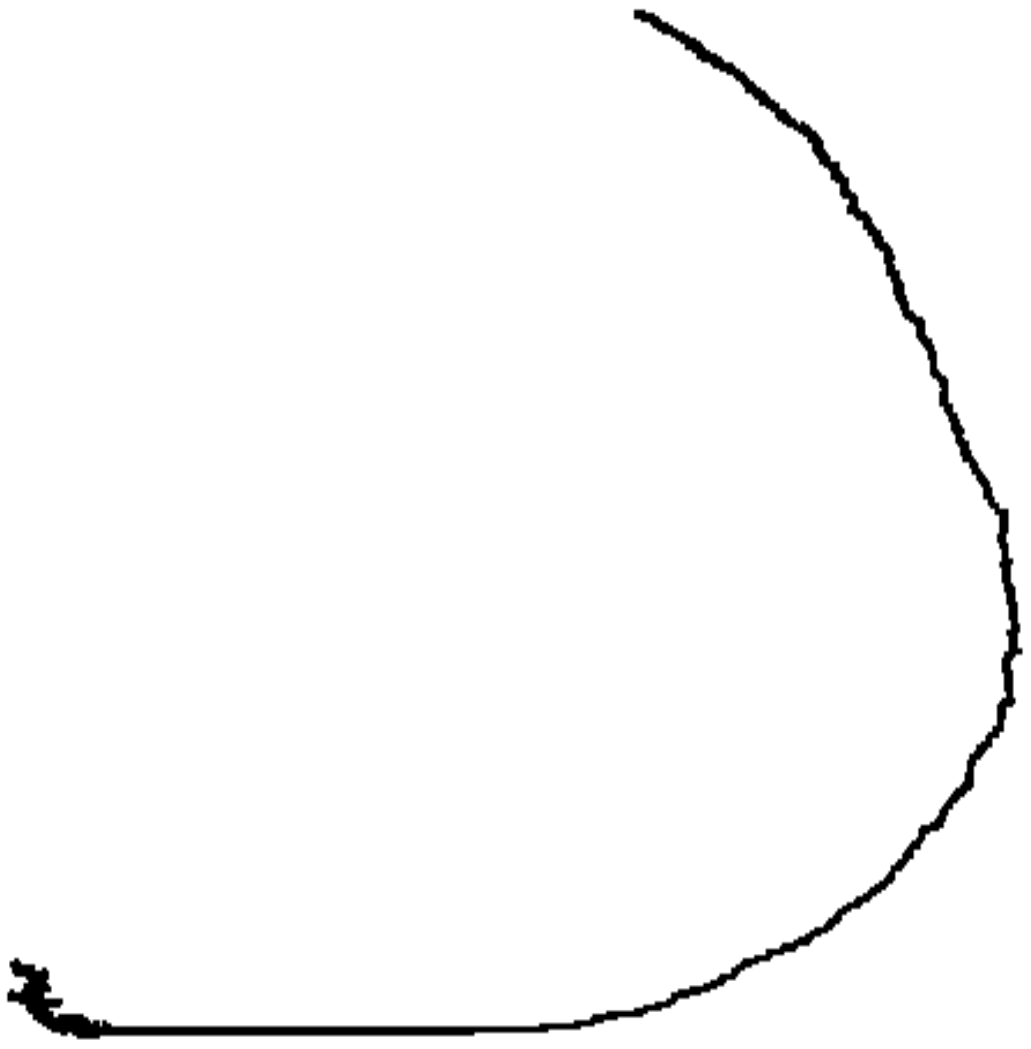

**Fig. 9.** DLA cluster with rotation: $\omega = 4.5 \cdot 10^{-3}$, M = 5000

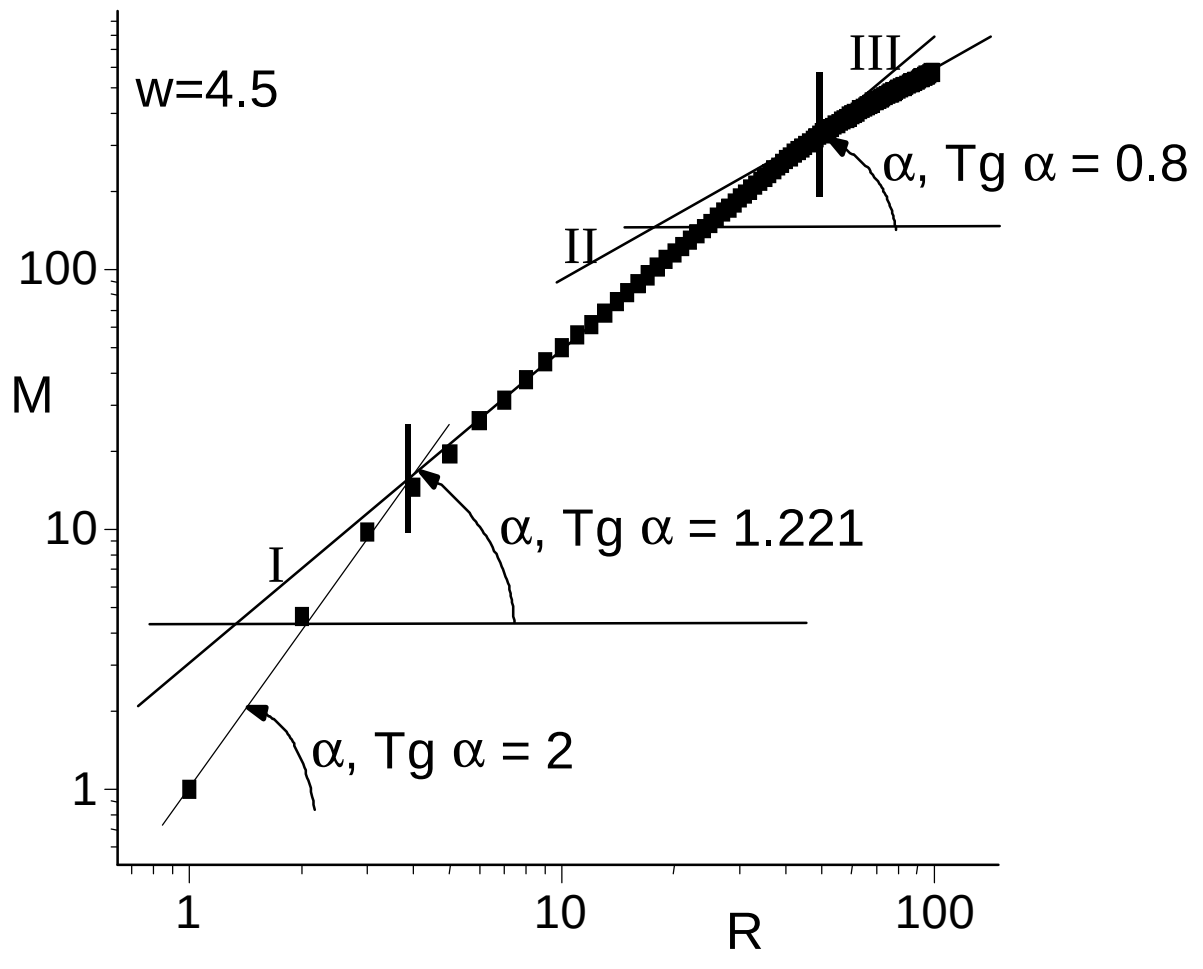

**Fig. 10.** Mass vs. Radius dependence for rotating DLA cluster: $\omega = 4.5 \cdot 10^{-3}$

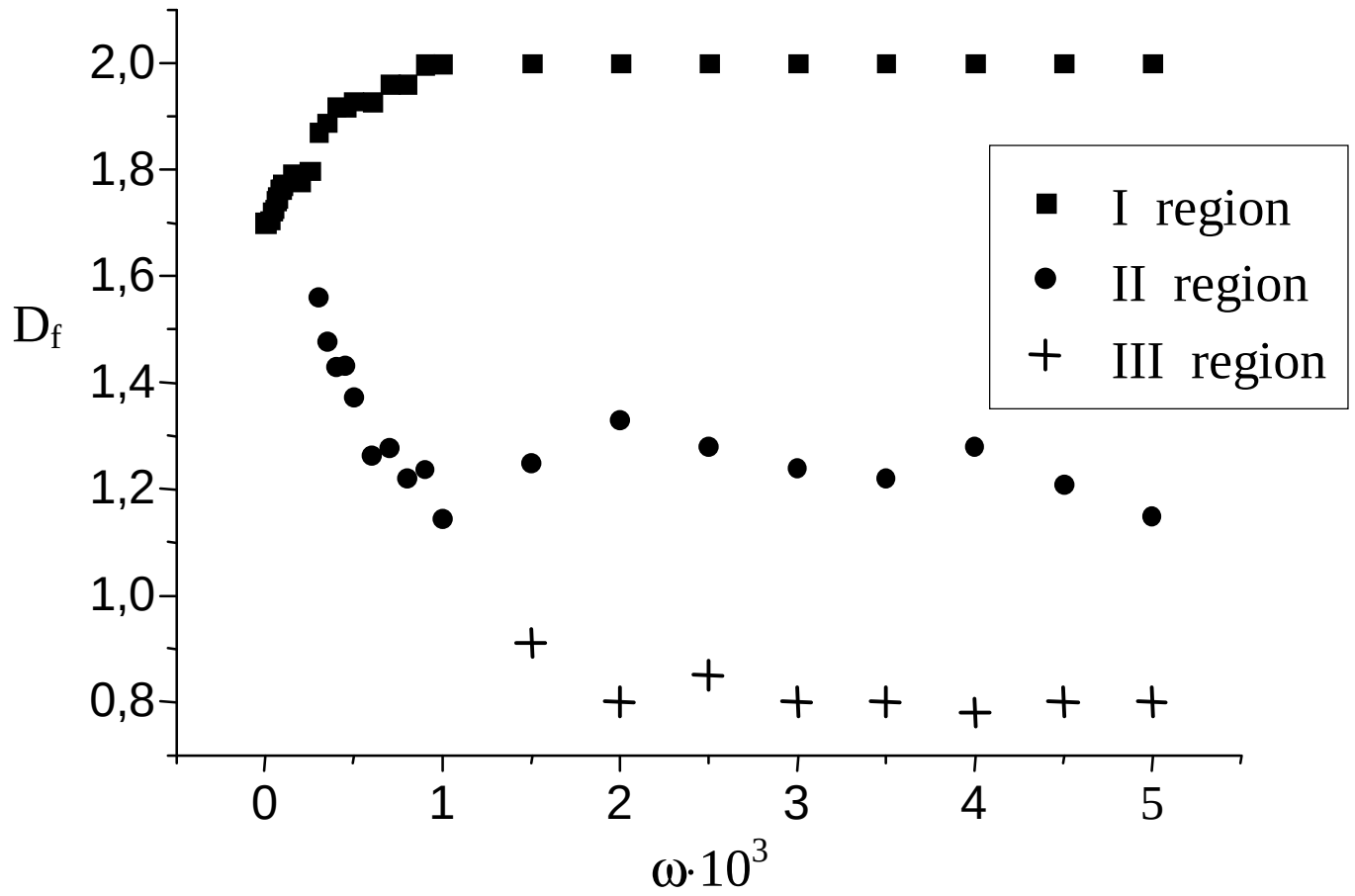

**Fig. 11.** Dependence of fractal dimension on rotation velocity

## Concluding Remarks

As our main result, we report the fractal dimension $D_f$ as a function of the angular velocity for two-dimensional DLA-cluster under rotation. This dependence has been obtained both by means of a new statistical annular model of the fractal growth and by means of direct computer simulation. The fractal dimension decreases when the rotation velocity is growing. We have found the clear evidence of a transition between fractal (at low rotation velocities) and non-fractal (at high velocities) regimes. The DLA-cluster under rotation shows different fractal dimensions when analyzed on different length scales. To analyze the multifractal properties of DLA-cluster in more detail the growth-site probability density should be calculated. This is what we intend to do in the nearest future.

## Acknowledgements

We would like to thank Prof. S. Nechaev for useful discussions.